\documentclass[twocolumn,showpacs,preprintnumbers,amsmath,amssymb]{revtex4}

\usepackage{amsmath,amstext,amsgen,amsbsy,amsopn,amsfonts,graphicx,amssymb}

\begin{document}

\title{Modulated Amplitude Waves in Bose-Einstein Condensates}

\author{Mason A. Porter}

\email[]{mason@math.gatech.edu}

\homepage[]{www.math.gatech.edu/~mason}

\affiliation{School of Mathematics and Center for Nonlinear Science, Georgia 
Institute of Technology, Atlanta, GA  30332}

\author{Predrag Cvitanovi\'c}

\affiliation{Center for Nonlinear Science and School of Physics, Georgia 
Institute of Technology, Atlanta, GA  30332}

\date{\today}

\begin{abstract}

	We analyze spatio-temporal structures in the Gross-Pitaevskii 
equation to study the dynamics of quasi-one-dimensional Bose-Einstein 
condensates (BECs) with mean-field interactions.  A coherent structure ansatz
 yields a parametrically forced nonlinear oscillator, to which we apply 
Lindstedt's method and multiple-scale perturbation theory to determine the 
dependence of the intensity of periodic orbits (``modulated amplitude 
waves'') on their wave number.  We explore BEC band structure in detail using
 Hamiltonian perturbation theory and supporting numerical simulations. 

\end{abstract}



\maketitle

	At low temperatures, particles in a gas can reside in the same 
quantum (ground) state, forming a Bose-Einstein 
condensate~\cite{pethick,stringari,ketter,edwards,becrub,becna}.  When 
considering only two-body, mean-field interactions, the condensate 
wavefunction $\psi$ satisfies the Gross-Pitaevskii (GP) equation, a cubic 
nonlinear Schr\"odinger equation (NLS) with an external potential:
\begin{equation}
	i\hbar\psi_t = -[\hbar^2/(2m)]\psi_{xx} + g|\psi|^2\psi + V(x)\psi
\,, \label{nls3}
\end{equation}
where $m$ is the mass of a gas particle, $V(x)$ is an external 
potential, $g = [4\pi\hbar^2 a/m][1 + \mathcal{O}(\iota^2)]$, $a$ is the 
(two-body) $s$-wave scattering length and $\iota = \sqrt{|\psi|^2|a|^3}$ is 
the dilute gas parameter~\cite{stringari,kohler}.  The quantity $a$ is 
determined by the atomic species in the condensate.  Interactions between 
atoms are repulsive when $a > 0$ and attractive when $a < 0$.  When 
$a \approx 0$, one is in the ideal gas regime.

 The quasi-one-dimensional (quasi-1d) regime employed in (\ref{nls3}) is 
suitable when the transverse dimensions of the condensate are on the order of
 its healing length and its longitudinal dimension is much larger than its 
transverse ones~\cite{bronski,bronskirep,bronskiatt,stringari}.  In this 
situation, one employs the 1d limit of a 3d mean-field theory rather than a 
true 1d mean-field theory, as would be appropriate were the tranverse 
dimension on the order of the atomic interaction length or the atomic size. 

In this paper, we examine uniformly propagating coherent structures by 
applying the ansatz $\psi(x-vt,t) = R(x - vt)\exp\left(i\left[\theta(x-vt) - \omega t\right]\right)$, where $R \equiv |\psi|$ is the magnitude (intensity) 
of the wavefunction, $v$ is the velocity of the coherent structure, 
$\theta(x)$ determines its phase, and $\omega$ is the temporal frequency 
(chemical potential).  Considering a coordinate system that travels with 
speed $v$ (by defining $x' = x - vt$ and relabeling $x'$ as $x$) yields
\begin{equation}
	\psi(x,t) = R(x)\exp\left(i\left[\theta(x) - \omega t\right]\right)
\,.  \label{maw2}
\end{equation}
When the (temporally periodic) coherent structure (\ref{maw2}) is also 
spatially periodic, it is called a {\it modulated amplitude wave} (MAW).  The 
orbital stability of coherent structures (\ref{maw2}) for the GP with 
elliptic potentials has been studied by Bronski and 
co-authors~\cite{bronski,bronskiatt,bronskirep}.  To obtain information about 
sinusoidal potentials, one takes the limit as the elliptic modulus $k$ 
approaches zero~\cite{lawden,rand}.  When $V(x)$ is periodic, the resulting 
MAWs generalize the Bloch modes that occur in linear systems with periodic 
potentials, as one is considering a nonlinear Floquet-Bloch theory rather 
than a linear one~\cite{675,ashcroft,band,space1,space2}.  

In this paper, we employ phase space methods and Hamiltonian perturbation 
theory to examine the band structure of such MAWs.  Prior work in this area 
has utilized numerical simulations~\cite{band,space1,space2}.

The novelty of our work lies in its illumination of BEC band structure 
through the use of perturbation theory and supporting numerical simulations 
to examine $2n\!:\!1$ spatial subharmonic resonances in BECs in period 
lattices.  Such resonances correspond to spatially periodic solutions of 
period $2n$ and generalize the `period doubled' states studied by 
Machholm, {\it et. al.}\cite{pethick2} and observed experimentally by 
Cataliotti, {\it et. al.}.\cite{cata}  

	Inserting (\ref{maw2}) into the GP (\ref{nls3}), equating 
real and imaginary parts, and simplifying yields
\begin{align}
	R' &= S\,, \notag \\
	S' &= \frac{c^2}{R^3} - \frac{2m\omega R}{\hbar} 
+ \frac{2mg}{\hbar^2}R^3 + \frac{2m}{\hbar}V(x)R\,. \label{dynam35}
\end{align}
The parameter $c$ is defined via the relation $\theta'(x) = c/R^2$, which is 
an expression of conservation of angular momentum~\cite{bronski}.  Null 
angular momentum solutions, which constitute an important special case, 
satisfy $c = 0$.

\begin{figure}[htb] 
	\begin{centering}
		\leavevmode
		\includegraphics[width = 2.5 in, height = 3 in,clip=]{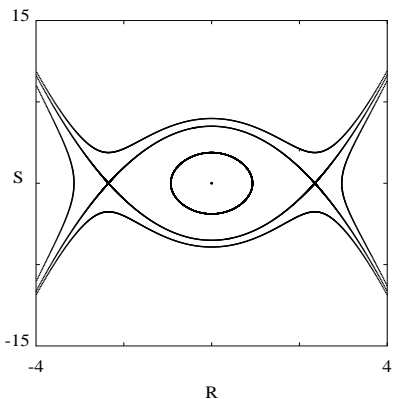}

	\vspace{ -.4 in}

		\caption{Phase portrait of a repulsive 
BEC with no external potential and $\omega = 10$.  In this plot, the two-body
 scattering length is $a = 0.072$ nm, obtained for atomic hydrogen 
($^1H$)~\cite{fried}.  Orbits inside the separatrix have bounded intensity 
$R(x)$ with increasing period as one approaches the separatrix.  The 
quantities $R$ and $S = R'$ are scaled quantities obtained with $m = 1/2$ and
 $\hbar = 1$.} \label{repulse1}
	\end{centering}
\end{figure}

	When $V(x) \equiv 0$, the two-dimensional dynamical system 
(\ref{dynam35}) is autonomous and hence integrable.  Its equilibria and the 
stability thereof are discussed in detail in Ref.~\cite{mapbec}.  When 
$g >0$, $\omega > 0$, and $c = 0$, which is the primary case for which we 
study the band structure, one obtains a neutrally stable equilibrium 
(a center) at $(R,S) = (0,0)$ and unstable equilibria (saddles) at 
$(\pm \sqrt{\hbar\omega/g},0)$.  See Fig.~\ref{repulse1}.

We employed Lindstedt's method to study the dependence of the wave number of 
periodic orbits (centered at the origin) of (\ref{dynam35}) on the intensity 
$R$ when $V(x) \equiv 0$~\cite{675}.  We assumed $g = \varepsilon \bar{g}$, 
where $\varepsilon \ll 1$ and $\bar{g} = \mathcal{O}(1)$.  The wave number is 
then $\alpha = 1 - 3gA^2/(8\omega\hbar) + \mathcal{O}(\varepsilon^2)$, where 
$R(\xi) = R_0(\xi) + \mathcal{O}(\varepsilon)$, $\xi := \alpha x$, 
$R_0(\xi) = A\cos(\beta\xi)$, and $\beta := \sqrt{2m\omega/\hbar}$.

\begin{figure}[htb] 
	\begin{centering}
		\leavevmode
		\includegraphics[width = 2.5 in, height = 3 in]{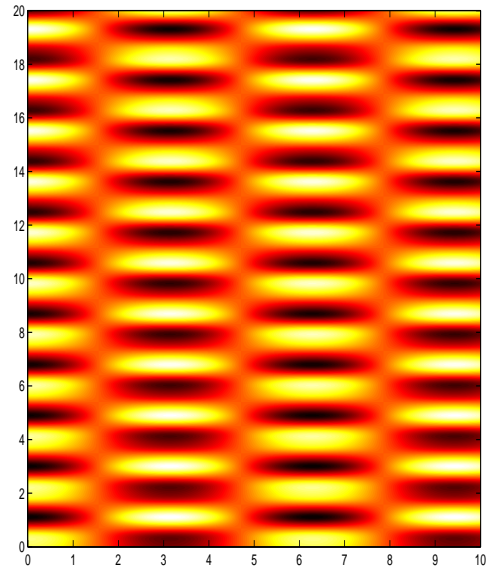}

	\vspace{-.1 in}

		\caption{An example of a spatially quasiperiodic coherent 
structure for $^1H$ in a sinusoidal lattice.   As in Fig.~\ref{repulse1}, 
$\omega = 10$, $a = 0.072$, $c = 0$, $m = 1/2$, and $\hbar = 1$.  
Additionally, $V_0 = 10$.  This plot, which was obtained from the coherent 
structure ansatz, depicts $\mbox{Re}(\psi)$, with the initial point 
$(R(0),S(0)) = (0.05, 0.05)$ inside the separatrix in Fig.~\ref{repulse1}.  
The darkest portions are the most negative, and the lightest are the most 
positive.  The quantities $R$ and $S$ are defined as in Fig.~\ref{repulse1}.}
 \label{mawh1010}
	\end{centering}
\end{figure}

To study the wave number-intensity relations of periodic orbits in the 
presence of external potentials, we expand the spatial variable $x$ in 
multiple scales.  We define ``stretched space'' $\xi := \alpha x$ as in the 
integrable situation and ``slow space'' $\eta := \varepsilon x$.  We consider 
potentials of the form $V(x) = \varepsilon \bar{V}(\xi,\eta)$, where 
$\bar{V}(\xi,\eta) = \bar{V}_0\sin\left[\kappa(\xi-\xi_0)\right] 
+ \bar{V}_1(\eta)$ and $\bar{V}_1(\eta)$, which is of order $\mathcal{O}(1)$,
 is arbitrary but slowly varying.  Cases of particular interest include 
$\bar{V}_1(\eta) = 0$ (periodic potential) and 
$\bar{V}_1(\eta) = \bar{V}_h(\eta - \eta_0)^2$ (superposition of periodic and 
harmonic potentials).  When $\bar{V}_h \ll \bar{V}_0$, this latter potential 
is dominated by its periodic contribution for many 
periods~\cite{kutz,promislow}.  The wavenumber parameter is 
$\kappa = 2\pi/T$, where $T$ represents the periodicity of the underlying 
lattice.  Optical lattices with more than twenty periods have now been 
created experimentally.\cite{lattice}  Spatially periodic potentials, which 
is the primary case we consider, have been 
employed in experimental studies of BECs~\cite{hagley,anderson}.  They have 
also been studied theoretically in Refs.~\cite{bronski,bronskiatt,bronskirep,space1,space2,promislow,kutz,malopt}.  An example of a coherent structure for 
hydrogen in the presence of a periodic lattice is depicted in 
Fig.~\ref{mawh1010}.  Coherent structures in other situations, such as for 
$^{85}Rb$ (for which $a = -0.9$), are examined in Ref.~\cite{mapbec}.

When $\kappa \neq \pm 2\beta$, the wave number-intensity relation for 
periodic orbits of (\ref{dynam35}) is 
\begin{equation}
	\alpha(C) = 1 - \frac{3g}{8\omega\hbar}C^2 
- \frac{1}{2\omega}V_1(x) + \mathcal{O}(\varepsilon^2)\,,  \label{alf35}
\end{equation}
where $R_0(\xi,\eta) = A(\eta)\cos(\beta\xi) + B(\eta)\sin(\beta\xi)$ and 
$C^2 = A^2 + B^2$ is a constant.  (Note that in equation (\ref{alf35}) and in
 our forthcoming discussion, the small paramater $\varepsilon$ has been 
absorbed back into the constants, so that we need not utilize bars over these
 quantities.)  When $\kappa = \pm 2\beta$, one obtains an 
extra term due to $2\!:\!1$ resonance:
\begin{equation}
	\alpha_R(C) = \alpha(C) 
\mp \frac{V_0}{4\omega} + \mathcal{O}(\varepsilon^2)\,, \label{perper35}
\end{equation}
where $C$ is defined as before but is no longer constant, and the sign of 
$V_0/4\omega$ alternates depending on which equilibrium of the slow dynamics 
of (\ref{dynam35}) one is considering~\cite{mapbec}.  Note that the parameter
 $\varepsilon$ has been absorbed back into the external potential in 
equations (\ref{alf35}) and (\ref{perper35}).

To examine the band structure of BECs in periodic lattices, we expand 
(\ref{dynam35}) with $V(x) \equiv 0$, $c = 0$ in terms of its exact elliptic 
function solutions, convert to action-angle variables, and apply several 
canonical transformations to obtain a ``resonance Hamiltonian,'' which we 
study both analytically and numerically.  Using elliptic functions rather 
than trigonometric functions allows one to analyze $2n\!:\!1$ subharmonic 
resonances with a leading-order perturbation expansion~\cite{mapbec,zounes}.  
We focus here on the case $g > 0$, $\omega > 0$.  In Ref.~\cite{mapbec}, we
 also discuss the implication of the work of Zounes and Rand~\cite{zounes} for
 the case $g < 0$, $\omega > 0$ and briefly consider the technically more 
complicated case $g < 0$, $\omega < 0$.  Refs.~\cite{band,space1,space2} 
concentrated on numerical studies of band structure.  In contrast, we employ 
Hamiltonian perturbation theory and study the band structure of BECs in 
periodic potentials both analytically and numerically.

Let $\xi_0 = \pi/(2\kappa)$ and $V_1 (x) \equiv 0$, so that 
$V(x) = (2m/\hbar)V_0\cos(\kappa x)$.  Equation (\ref{dynam35}) is then 
written $R'' + \delta R + \alpha R^3 + \epsilon R \cos(\kappa x) = 0$, where 
$\delta = 2m\omega/\hbar > 0$, $\alpha = -2mg/\hbar^2 < 0$, and 
$\epsilon = -(2m/\hbar)V_0$.  (Note that the perturbation parameter 
$\epsilon$ is not the same as the parameter $\varepsilon$ employed earlier.) 
 When $V_0 = 0$, one obtains the exact elliptic function solution
\begin{equation}
	R = \mu \rho \, \mbox{cn}(u,k) \,, \label{ell0}
\end{equation}
where $u = u_1x + u_0$, $u_1^2 = \delta + \alpha \rho^2$, 
$k^2 = (\alpha \rho^2)/(2[\delta + \alpha \rho^2])$, $u_1 \geq 0$, 
$\rho \geq 0$, $k^2 \in \mathbb{R}$, and $\mu \in \{-1,1\}$.  The initial 
condition parameter $u_0$ can be set to $0$ without loss of generality.  We 
consider $u_1 \in \mathbb{R}$ in order to study periodic solutions inside the
 separatrix (depicted in Fig.~\ref{repulse1}).  One need not retain the 
parameter $\mu$ to do this, so we set it to unity.  Because 
$k^2 \in [-\infty,0]$, we utilized the reciprocal complementary modulus 
transformation in deriving (\ref{ell0})~\cite{coppola,copthes,stegun,mapbec}.
  Defining $\chi = \sqrt{\delta}x$ and $r = \sqrt{-\delta/\alpha}R$ and 
denoting $' := d/d\chi$, the equations of motion take the form
\begin{equation}
	r'' + r - r^3 + \epsilon \delta^{-1}\cos\left(\kappa \delta^{-1/2} \chi\right)r = 0 \,, \label{dynamo}
\end{equation} 
with the corresponding Hamiltonian (using $s := r'$)
\begin{equation}
	H(r,s,\chi) = \frac{1}{2}s^2 + \frac{1}{2}r^2 - \frac{1}{4}r^4 + 
\frac{\epsilon}{2\delta}r^2\cos\left(\frac{\kappa}{\sqrt{\delta}}\chi\right)
\,. \label{hamham}
\end{equation}

The frequency of a given periodic orbit is 
$\Omega(k) = \pi\sqrt{1-\rho^2}/(2K(k))$, where $K(k)$ is the complete 
elliptic integral of the first kind~\cite{watson}.  One thereby obtains the 
action~\cite{goldstein,gucken,wiggins,lich}
\begin{equation}
	J = \frac{4\sqrt{1-\rho^2}}{3\pi}\left[E(k) 
- \left(1 - \frac{\rho^2}{2}\right)K(k)\right] \,, \label{action}
\end{equation}
where $E(k)$ is the complete elliptic integral of the second kind, and the 
conjugate angle $\Phi := \Phi(0) + \Omega(k)\chi$.  The frequency $\Omega(k)$
 decreases monotonically as $k^2$ goes from $-\infty$ to $0$ (as one goes 
from the separatrix to $(0,0)$).  

After applying several near-identity canonical transformations and expanding 
elliptic functions in Fourier series~\cite{mapbec}, one obtains an autonomous 
resonance Hamiltonian $K_{n}(Y,\xi)$ (in action-angle coordinates) for the 
$2n\!:\!1$ resonance band,
\begin{equation}
	K_{n} = Y - Y^2 - \frac{\kappa J(Y)}{2n\sqrt{\delta}} + \frac{\epsilon}{2\delta}Y\mathcal{B}_{n}(Y)\cos\left(\frac{2n\xi}{J'(Y)}\right) \,, 
\label{ram}
\end{equation}
where $\mathcal{B}_{n}$ are obtained from Fourier 
coefficients~\cite{mapbec,zounes}.  The band associated with $2n\!:\!1$ 
subharmonic spatial resonances is present when 
$\kappa/\sqrt{\delta} \leq 2n$.
 
To obtain an analytical description of these resonance bands, we note that 
such bands emerge from the action $Y = Y_{n}$, the location of the $n$th 
resonance torus in phase space, which is determined by 
$\kappa/\sqrt{\delta} = 2n\Omega(Y_{n})$.  The saddles and centers of this 
resonance band are given, respectively, by $Y_s = Y_{n} - |\Delta Y|$ and 
$Y_c = Y_{n} + |\Delta Y|$,
\begin{equation}
	\Delta Y = \mp\frac{\epsilon}{2\delta}\left[\frac{\mathcal{B}_{n}(Y_{n}) + Y_{n}\mathcal{B}_{n}'(Y_{n})}{\Omega(Y_{n})\sqrt{1-2Y_{n}}\tilde{K}'(Y_{n}) - 1}\right] \,,
\end{equation} 
where $\tilde{K}'(Y) := 2K'(k(Y))/\pi$, $K'(k) := K(\sqrt{1-k^2})$, 
$\Delta Y > 0$ when $n$ is even, and $\Delta Y < 0$ when $n$ is odd.  The 
width of resonance bands is
\begin{equation}
	W = 2\left | \epsilon\frac{Y_{n}\mathcal{B}_{n}(Y_{n})}{\delta\left[1 + \frac{\kappa}{2n\sqrt{\delta}}J''(Y_{n})\right]}\right |^{1/2} \,.
\end{equation}

\begin{figure}[htb] 
	\begin{centering}
		\leavevmode
		\includegraphics[width = 2.5 in, height = 3 in]{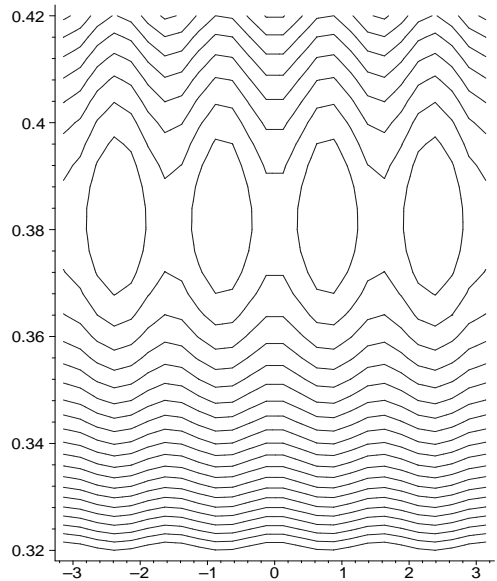}

\vspace{-.1 in}

		\caption{Resonance Hamiltonian $K_2$ for 
$\kappa = 2.5$, $\delta = 1$, and $\epsilon = 0.01$.  The vertical axis is in
 units of action $Y$, and the horizontal axis is in units of $\xi/J'(Y)$.  As
 discussed in the text, these quantities are scaled and physically unitless.} 
\label{k201}
	\end{centering}
\end{figure}

\begin{figure}[htb] 
	\begin{centering}
		\leavevmode
		\includegraphics[width = 2.5 in, height = 3 in]
{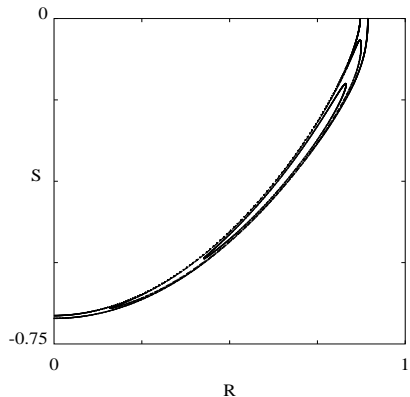}

\vspace{-.4 in}

		\caption{Lower right corner of a Poincar\'e section for 
$\kappa = 2.5$, $\delta = 1$, and $\epsilon = 0.01$.  Note that there is no 
$2\!:\!1$ resonance band for this choice of $(\kappa,\delta)$.  The 
$4\!:\!1$ resonance is depicted.  (Three additional copies of this structure 
appear in the Poincar\'e section.)  Recall that the scaled quantities $R$ and
 $S$ are unitless.} \label{res2}
	\end{centering}
\end{figure}

We compare our analytical results with numerical simulations in $(R,S)$ 
coordinates with $\alpha = -1$, $m = 1/2$, and $\hbar = 1$.  For example, 
when $\kappa = 2.5$ and $\delta = 1$, the $4\!:\!1$ resonances are the 
lowest-order resonances present.  The resonance Hamiltonian $K_2$ is depicted
 in Fig.~\ref{k201} for $\epsilon = 0.01$, and the corresponding Poincar\'e 
section is shown in Fig.~\ref{res2}.  From (\ref{ram}), one predicts that the
 $R$-axis saddles are located at $(R,S) = (\pm 0.86364,0)$, which is rather 
close to the true value of about $(\pm 0.88,0)$.  The $S$-axis saddles are 
predicted to occur at $(R,S) = (0, \pm 0.68389)$, whereas the true value is 
about $(0, \pm 0.687)$.  Numerous other examples are studied in 
Ref.~\cite{mapbec}.

At the center of the KAM islands, we observe `period-multiplied' states.  
When $n = 1$, these correspond to the period-doubled states studied recently
 by Machholm and coauthors.\cite{pethick2}

We note, finally, that to analyze three-body interactions (which is 
necessary, for example, to examine Feshbach resonances~\cite{fesh}), one has 
to take dissipative effects into account~\cite{kohler,rub85}.  In the present 
paper, we studied only two-body interactions.

In sum, we employed Lindstedt's method and multiple scale analysis to 
establish wave number-intensity relations for MAWs of BECs in periodic 
lattices.  With this approach, we studied $2\!:\!1$ spatial resonances and 
illustrated the utility of phase space analysis and perturbation theory for 
the study of band structure as well as the structure of modulated amplitude 
waves in BECs.  Using a more technically demanding perturbative approach 
relying on the elliptic function structure of solutions of the integrable GP 
yields the $2n\!:\!1$ spatial resonances (band structure) of MAWs, which we 
studied in considerable detail both analytically and numerically.

Valuable conversations with Eric Braaten, Michael Chapman, Mark Edwards, 
Nicolas Garnier, Brian Kennedy, Panos Kevrekidis, Yueheng Lan, Boris Malomed,
 Igor Mezi\'c, Peter Mucha, and Dan Stamper-Kurn are gratefully acknowledged.
  We are especially grateful to Jared Bronski, Richard Rand, and Li You for 
several extensive discourses.

\bibliographystyle{plain}

\end{document}